# Experimental observation of the Aubry transition in two-dimensional colloidal monolayers

T. Brazda[1], A. Silva[2], N. Manini[2], A. Vanossi[3,4], R. Guerra[2,5], E. Tosatti[4,6,3], and C. Bechinger[1,7*]

[1]2. Physikalisches Institut, Universität Stuttgart, Pfaffenwaldring 57, 70569 Stuttgart, Germany

[2]Dipartimento di Fisica, Università degli Studi di Milano, Via Celoria 16, 20133 Milano, Italy

[3]CNR-IOM Democritos National Simulation Center, Via Bonomea 265, 34136 Trieste, Italy

[4]International School for Advanced Studies (SISSA), Via Bonomea 265, 34136 Trieste, Italy

[5]Center for Complexity and Biosystems, University of Milan, 20133 Milan, Italy,

[6]International Centre for Theoretical Physics (ICTP), Strada Costiera 11, 34014 Trieste, Italy

[7]Fachbereich Physik, Universität Konstanz, Universitätsstr. 10, 78464 Konstanz, Germany

*Correspondence to: clemens.bechinger@uni-konstanz.de

The possibility to achieve entirely frictionless, i.e. superlubric, sliding between solids, holds enormous potential for the operation of mechanical devices. At small length scales, where mechanical contacts are well-defined, Aubry predicted a transition from a superlubric to a pinned state when the mechanical load is increased. Evidence for this intriguing Aubry transition (AT), which should occur in one dimension (1D) and at zero temperature, was recently obtained in few-atom chains. Here, we experimentally and theoretically demonstrate the occurrence of the AT in an extended two-dimensional (2D) system at room temperature using a colloidal monolayer on an optical lattice. Unlike the continuous nature of the AT in 1D, we observe a first-order transition in 2D leading to a coexistence regime of pinned and unpinned areas. Our data demonstrate that the original concept of Aubry does not only survive in 2D but is relevant for the design of nanoscopic machines and devices at ambient temperature.

In the expanding fields of nanoscience, where the competition of length scales is of key importance, Aubry's theoretical concept [1, 2] how to achieve frictionless sliding, is one of the most challenging topics in nanotribology [3], with immediate technological applications. The nowadays pervasive tribological concept of superlubricity was originally inferred from the one-dimensional (1D) Frenkel-Kontorova (FK) class of models [4], describing a chain of interacting particles subject to a periodic substrate potential. Aubry showed that, in the thermodynamic limit and at zero temperature, the 1D incommensurate chain/substrate interface may undergo a second-order transition between an unpinned and a pinned state at a critical value of the substrate corrugation, i.e. contact strength, $U_0$. In the unpinned state, the minimum force required to achieve sliding, i.e. the static friction, should vanish: the interface incommensurability can indeed prevent asperity interlocking and collective stick-slip motion of the interface atoms, with a consequent negligibly small frictional force.

After remaining locked away for several decades, Aubry-type signatures were recently observed experimentally in finite 1D linear laser-cooled ion-chains on corrugated potentials [5-8]. In contrast, the demonstration and characterization of the Aubry transition in extended and mismatched two-dimensional (2D) monolayers at finite temperature, corresponding to more realistic conditions, is still lacking. Previous experimental work on 2D systems including telescopic dynamics [9] and water flow [10] in carbon nanotubes, sliding of graphite flakes [11-13] and mesas [14-16], graphene nanoribbons [17], cluster nanomanipulation [18, 19], or rare-gas island inertial motion [20] has demonstrated superlubricity, i.e. an ultra-low static friction, scaling sublinearly with the contact size. However, without the possibility of tuning interaction parameters, the nature of the Aubry transition at 2D interfaces cannot be unravelled and does not disclose how a change from a superlubric to a pinned state occurs.

Here, we report the first microscopically-resolved experimental study of the Aubry transition by investigating an extended colloidal layer driven over a laser-generated periodic potential. Unlike conventional substrates, here, all relevant physical parameters, such as the lattice periodicity and the contact strength can be controlled in situ [21]. Unlike 1D, a sliding 2D crystal will develop a misalignment angle relative to the optical lattice, being crucial for its frictional properties. Our results demonstrate, that around the critical contact strength, the 2D Aubry transition brings about the occurrence of a novel phase separation with a coexistence region between pinned and free sliding colloidal patches. The frictional rheology of the monolayer is especially interesting and heterogeneous in this two-phase region, which we also characterise by means of molecular-dynamics (MD) simulations mimicking the experimental conditions.

Our experiments were performed at $T_0 \simeq 298$ K with a suspension of micron-sized polystyrene particles interacting via a repulsive screened Coulomb potential (see Methods in the Supplemental Material [22]). The colloids form a 2D hexagonal crystal with lattice constant $a_C$ at the bottom of the sample cell. A periodic substrate is created by interfering three partial laser beams, as sketched in Fig. 1a. This optical landscape provides a hexagonal substrate potential, Fig. 1b, whose amplitude $U_0$ can be tuned by the laser intensity. The laser lattice constant $a_L$ is set by the intersection angle of the laser beams and allows us to adjust the mismatch ratio $\varepsilon = a_L/a_C$ Particle positions are tracked by video microscopy with a resolution of about 50 nm (Fig.1b). A controlled driving force $F$ of the monolayer is exerted by viscous Stokes forces which arise when the sample cell is translated horizontally relative to the (resting) interference pattern [22].

Figure 1c shows experimental mobilities $\mu$ of an incommensurate ($\varepsilon = 0.84$) colloidal monolayer vs. $F$ and for different values of $U_0 = 17$ - $102\ k_B\ T_0$. For large $U_0$, the monolayer is strongly pinned to the substrate and remains immobile until $F$ exceeds the static friction force $F_s$, defined by the value where $\mu$ exceeds 10% of a free sliding (dashed line). Below $U_0 \leq 38\ k_B\ T_0$, however, the monolayer becomes mobile already at a minute driving force $F_{min} \simeq 1$ fN. This indicates a transition from a pinned to a superlubric state, in agreement with one of the signatures of the Aubry transition. Figure 1d shows the static friction force as a function of the corrugation amplitude. To obtain system-independent quantities $F_s$ is normalized to that of a single colloid in the same corrugation potential $F_{1s}$, and $U_0$ to the critical corrugation amplitude $U_c$ where the monolayer becomes pinned at $F_{min}$ (Fig. 1e). In agreement with the predictions by Aubry, our data show the disappearance of static friction below $U_0/U_c \simeq 1$. Above this transition, $F_s$ increases almost linearly and eventually saturates. The open symbols, which were obtained by numerical simulations with parameters adjusted to our experimental conditions [22], show excellent agreement with our data.

Because the Aubry transition is a structural phase transition, it must affect the monolayer microstructure. For stiff monolayers, minimization of the free energy is achieved by rotation of the lattice relative to the substrate by the Novaco angle $\theta_{Nov}$ [23-24]. Because most particles are slightly displaced from substrate minima after rotation, they become superlubric (unpinned). Contrary to very stiff monolayers, where $\theta_{Nov}$ has a sharp-defined value which depends on the mismatch ratio (for $\varepsilon = 0.84$ one obtains $\theta_{Nov} \simeq 5°$) [25], for deformable monolayers the misfit angle $\theta$ locally varies across the monolayer. Such behavior is confirmed by our experiments as seen by typical particle configurations for different substrate amplitudes (Fig. 2a-c). Red/blue colored particles belong to superlubric domains which are rotated by $|\theta| > ½\, \theta_{Nov} = 2.5°$, while grey particles correspond to pinned regions (Methods). With increasing $U_0$ the colloidal lattice becomes increasingly locally aligned with the substrate, as seen by the grey (aligned) domains which proliferate at the expense of tilted regions. For $U_0 = 11.3\ U_c$ (Fig.2c) essentially only pinned particles are observed. Figures 2d-f show how the misfit angle distribution gradually changes as a function of $U_0$. Simulations suggest that the Aubry transition in 2D is of first-order, which implies the coexistence of pinned and superlubric regions [26]. The structural features of the three phases are evident in the structure factor analysis reported in Figs S2 and S3 [22]. Upon ramping $U_0$ up and down, simulations exhibit a clear hysteresis in the fraction $f_{tilt}$ of superlubric particles (Fig. 2g). From the range where error bars show no overlap, the coexistence range between the pinned and unpinned particles can be estimated. Even if, unlike simulations, experimental data are sufficiently well equilibrated and do not show hysteresis, they nonetheless show clear evidence for coexistence (Fig.2b).

Within the 1D FK model a superlubric-pinned transition is structurally characterized by the displacement of all particles away from the maxima of the corrugation potential (Fig. 3a-c). Following the disorder parameter of Ref. [27,28], we evaluate the fraction $\Psi$ of colloids at positions where the local substrate potential is above that of the saddle points, namely the white triangles of Fig. 3d. This fraction should be 25% for randomly placed colloids or for $U_0$ = 0 and decreases with increasing $U_0$. This is in agreement with our experimental and simulation data (Fig. 3e-h). It should be emphasized that the $U_0$-dependence of Ψ is in good agreement with the corresponding $f_{tilt}$ behavior in Fig. 2. Both quantifiers support the existence of a first-order Aubry transition at $U_0/U_c \simeq 1$.

An important novelty in 2D compared to 1D is that depinning can occur inhomogeneously, nucleating at specific points in the sample. This is especially expected in the phase-coexistence region where, for a given applied force, superlubric patches tilted at the Novaco angle are likely to unpin earlier than non-tilted regions. To explore such behavior, we have simulated a configuration near the right end of the coexistence region ($U_0$ = 1.2 $U_c$). After applying a weak force $F < F_s$, $F$ is abruptly increased to $F$ = 1.09 $F_s$. The red line in Fig. 4a reports how the monolayer progressively depins in time. This progressive depinning is understood when examining the simulation movie (S.M.), with the local-angle labeling showing a progressive extension of the red/blue-colored unpinned regions. Figures 4b and 4c report two snapshots of that movie, representative of the mostly pinned initial and the mostly sliding steady state, respectively. Arrows showing individual particle displacements indicate that initial depinning occurs mainly at the interface regions between the two phases.

Our experimental approach provides a versatile way of studying the conditions under which superlubricity occurs in extended 2D contacts. In addition, we obtain previously unattainable insights into the microstructural changes occuring during the Aubry transition. For example, the observation of a first-order transition from a superlubric to a pinned state demonstrates that, in contrast to a smooth transition in 1D, as well as in an interesting model of a crystal grain boundary [29], the Aubry transition may remain sharp at finite temperatures for 2D contacts. The first-order character in our case is boosted by the local angular compliance coupled to the Novaco-McTague rotation. In a different 2D system lacking that compliance, such as e.g., a graphene flake on a crystal surface, the first order character is likely to be weakened. In particular for micro- and nano-mechanical devices whose contact areas are rather free from defects, asperities, and long-range elasticity effects [30-32], this should allow for drastic variations in the frictional behavior by tiny changes of the mechanical load.

## Acknowledgements

Work in Trieste was carried out under ERC Grant 320796 MODPHYSFRICT. The COST Action MP1303 is also gratefully acknowledged.

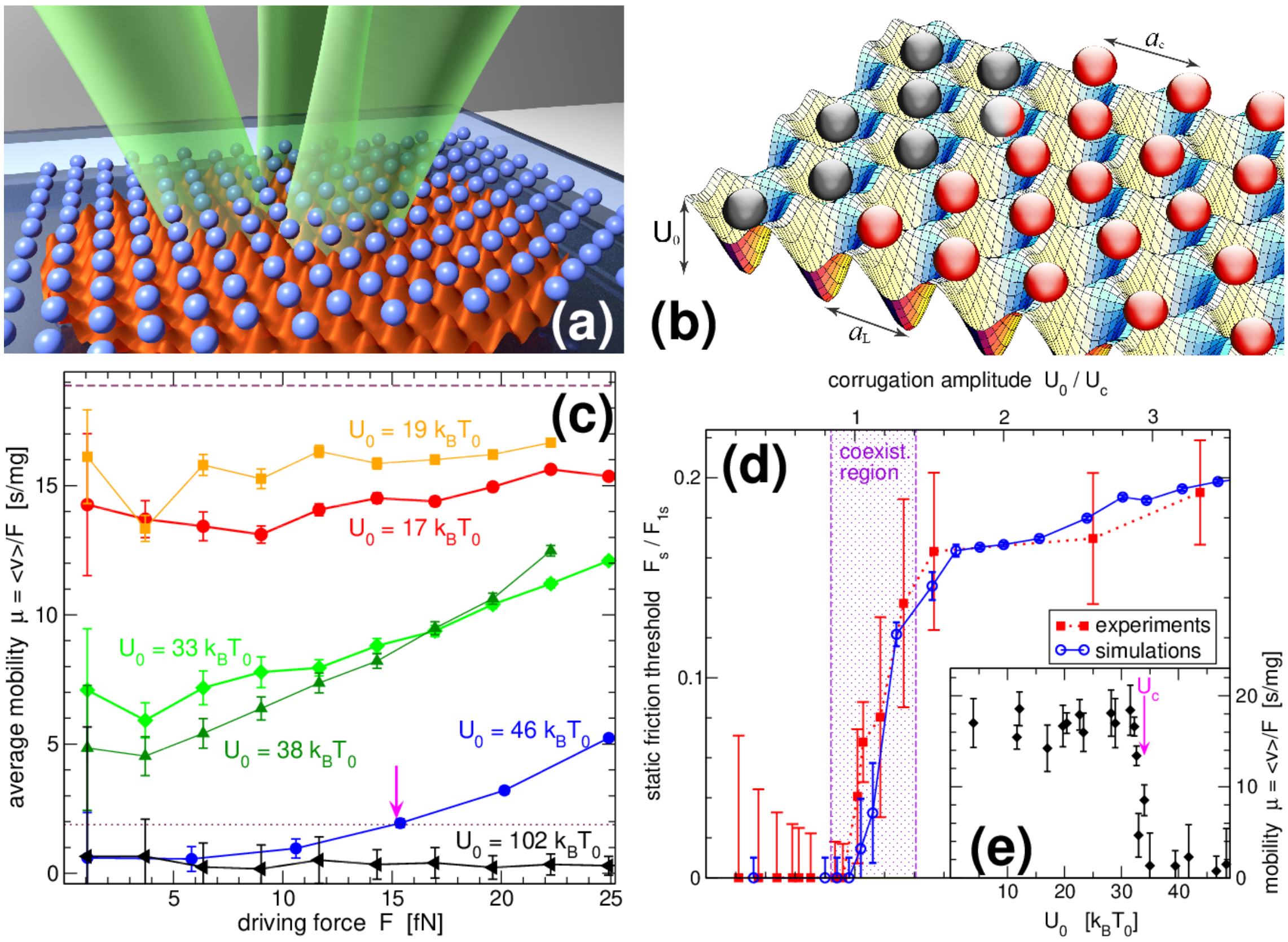

**Figure 1. Driving a colloidal monolayer across a periodic laser potential.** (a) Schematic view of the experimental setup. (b) Corresponding simulated model. (c) Measured mobility of an incommensurate (ε = 0.84) monolayer vs. driving force for different corrugation amplitudes $U_0$ in the superlubric (red, orange), coexistence (green), and statically pinned phase (blue, black); dashed line: the maximum mobility ($U_0$=0); dotted line: the minimum threshold mobility to detect sliding. (d) Static friction force $F_s$ vs. $U_0$ obtained from experiments (solid) and simulations (open); shaded area: the coexistence region, across which Aubry transition takes place. (e) The monolayer mobility under the action of the experimentally smallest accessible driving force $F_{min} \simeq 1$ fN; the critical corrugation $U_c = 34\ k_BT_0$ is defined by the sharp drop in the mobility (arrow).

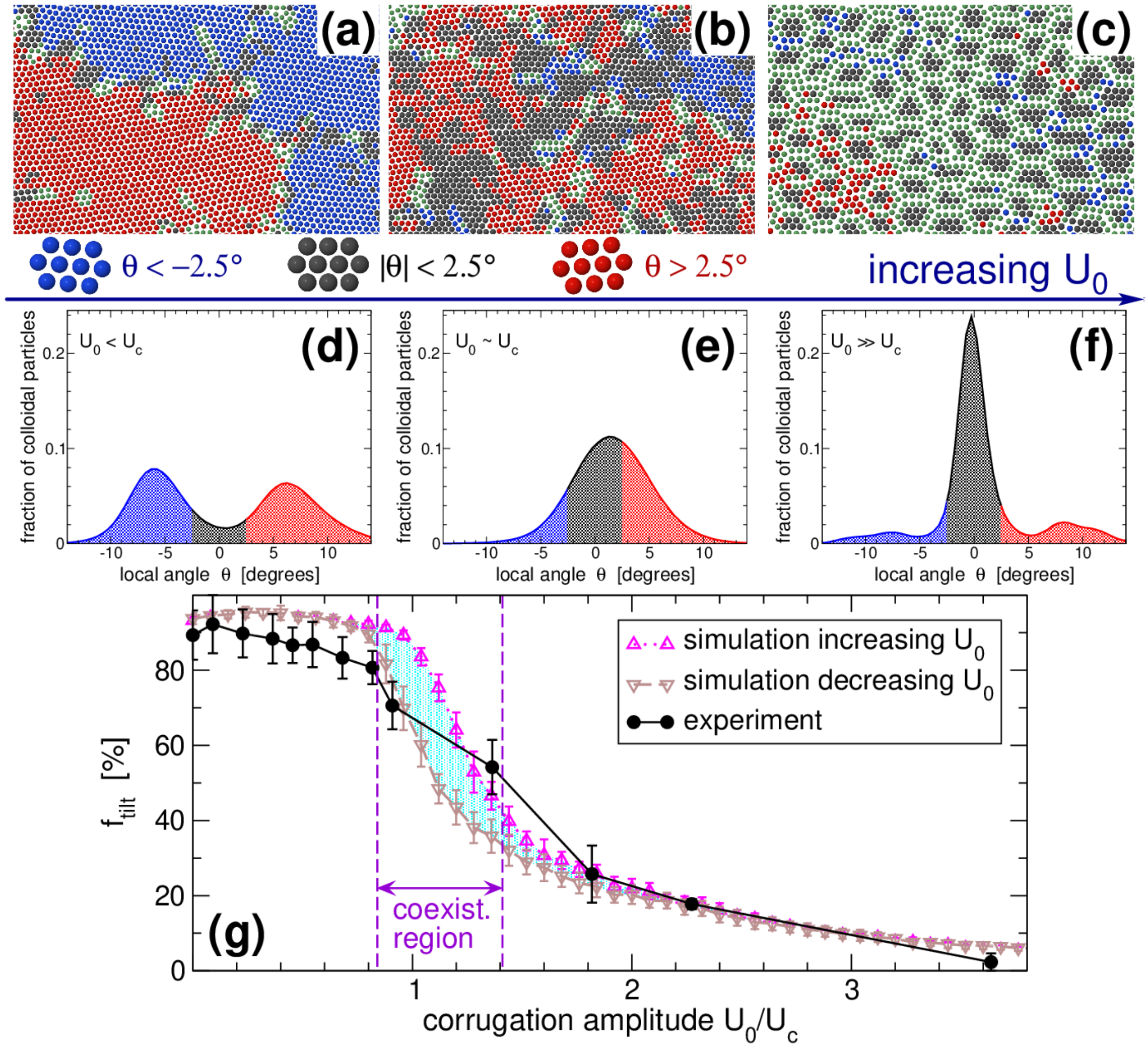

**Figure 2. Structural response and emergence of coexistence region.** (a-c) Configurations representative of (a) the superlubric phase, (b) the coexistence region, and (c) the pinned phase. The color scheme follows the local bond-orientation angle $\theta_i$ relative to the crystalline directions of the corrugation potential (S.M.). (d-f) The local-angle distribution, obtained for corrugation (d) $U_0 = 0.63\ U_c$ (unpinned/superlubric configuration), (e) $U_0 = 1.25\ U_c$ (coexistence region), and (f) $U_0 = 11.3\ U_c$ (statically pinned region). (g) Corrugation dependence of the fraction $f_{tilt}$ of tilted (red + blue) colloids. The presence of hysteresis, due to slow numerical equilibration, helps identifying in simulation the coexistence region, characterized by a change of slope in experiment.

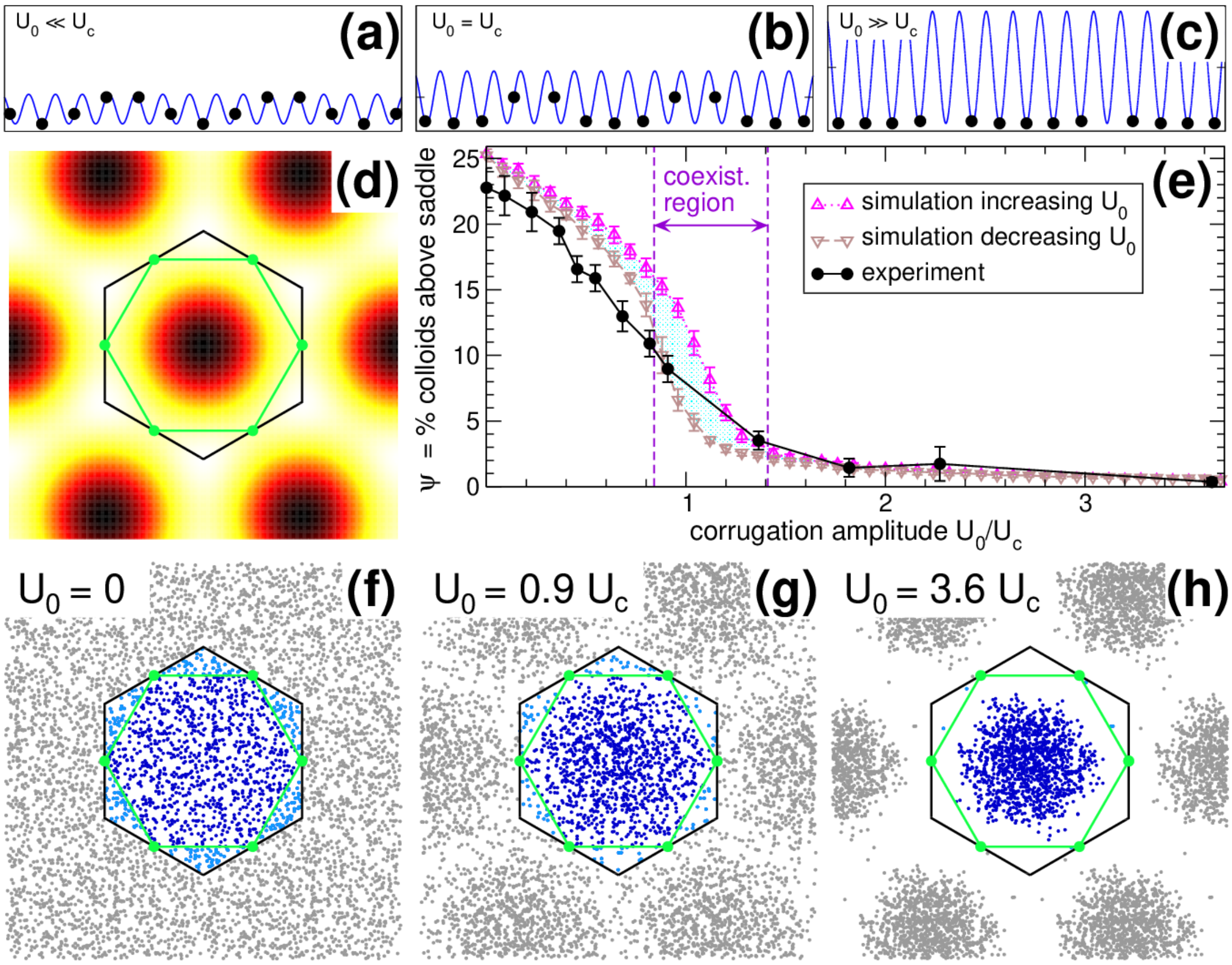

**Figure 3. Local energetics across the Aubry transition.** For increasing $U_0$ a substrate-mismatched 1D FK chain evolves from a configuration in which (a) minima and maxima regions are evenly populated (superlubric), (b) maxima regions start to depopulate (pinning threshold), and (c) only minima regions are populated (strongly pinned). (d) The 2D corrugation-potential profile (dark/light = low/high energy); the repulsive regions are comprised between the green (saddle-energy contour level) and the black (Wigner-Seitz cell) hexagons. (e) For the 2D colloid crystal, the fraction Ψ of colloids in the repulsive regions as a function of $U_0$ illustrating the statistical crossover from superlubric to pinned. (f), (g), (h) Experimental particle positions folded within the Wigner-Seitz cell of the corrugation lattice.

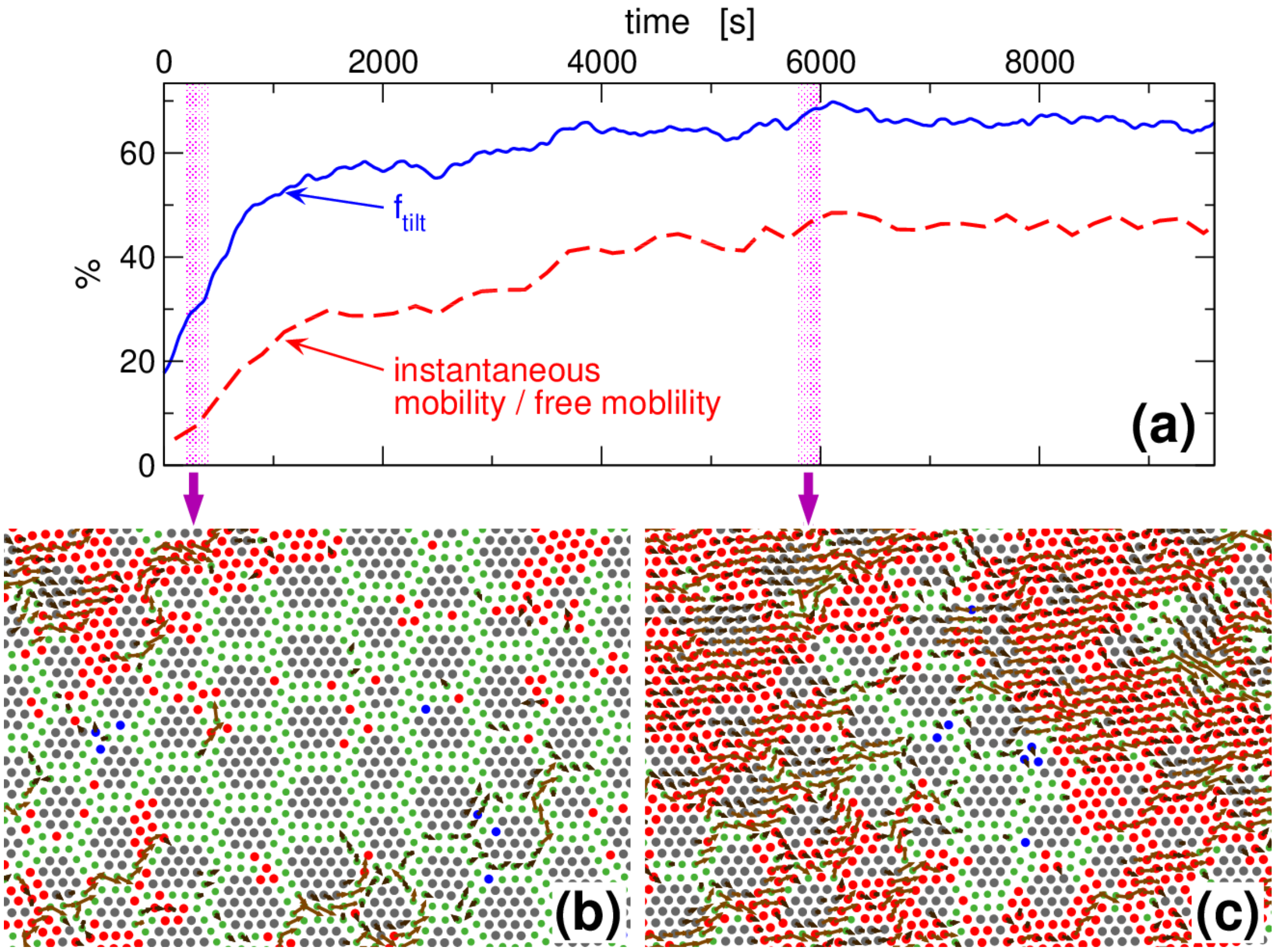

**Figure 4. Force-induced depinning transition.** In a simulated pinned configuration near the coexistence region ($U_0 = 1.2\ U_c$), the applied force is raised suddenly from $F = 0.73\ F_s$ to $F = 1.09\ F_s$, at time $t = 0$. (a) both $f_{tilt}$ and the overall mobility of the colloidal crystal increase steadily in time, until steady state. Starting from (b) a pinned phase ($t$ = 200 s), the applied force promotes the growth of the (c) superlubric tilted phase ($t$ = 5800 s). The colloids color scheme follows Fig. 2. Brown arrows indicate the displacement patterns over a 200 s time interval. See also the S.M. Movie.

# Supplemental Material

## 1. Methods

### 1.1 Experiments

We used carboxylated polystyrene particles with with radius $R$ = 1.95 µm. Due to their functionalization with –COOH groups, particles interact via a screened Coulomb potential

$$u(r) = A\frac{exp(-\kappa r)}{r}, \qquad \text{(S1)}$$

where $A$ depends on the colloidal surface charges and $\kappa^{-1}$ is the Debye screening length. Hexagonal optical interference patterns were created by overlapping three beams of equal intensity and polarization of a solid-state laser ($\lambda$ = 532 nm, $P$ = 18 W). An additional vertical incident laser beam with perpendicular polarization exerts a vertical optical pressure and pushes the particles towards the negatively charged silica bottom plate of the sample cell. This pressure reduces vertical position fluctuations to less than 5% of the particle diameter.

Lateral driving forces were applied by translation of the sample cell by a piezo-stage with velocity $v_d$ which leads to a lateral Stokes force on each particle $F = \eta_{eff}\, v_d$ where $\eta_{eff}$ is the effective drag coefficient $\eta_{eff} = 5.3 \times 10^{-8}$ kg s$^{-1}$, obtained from the short-time particle diffusion coefficient $D = \eta_{eff}\, k_B\, T$. To avoid that the driving force leads to inhomogeneities in the particle density, it was periodically reversed after about 1000 s. The smallest force is limited by the slowest velocity of the stage (20 nm/s) to $F_{min}$ = 1.06 ± 0.39 fN. The substrate corrugation of the sinusoidal light potential $U_0$ is measured by application of a lateral force to a dilute colloidal suspension. From the smallest force $F_{1s}$ which is required to depin a single colloid, $U_0$ can be evaluated according to $U_0 = 9\, a_L\, F_{1s}/(8\pi)$.

The mobility $\mu$ resulting from the application of a given driving force $F$ is obtained from the observed average advancement speed of the colloids $<v>$ by $\mu = <v>/F$. The largest possible mobility is the one obtained in the absence of corrugation ($U_0 = 0$), and equals $\mu_{max} = v_d/F = \eta_{eff}^{-1}$, dashed line in Fig. 1C.

In both simulations and experiments, the threshold mobility for the evaluation of the static friction $F_s$ is set at 10% of $\mu_{max}$, dotted line in Fig. 1C.

## 1.2 Simulations

The driven colloids are modelled as charged point particles undergoing overdamped 2D planar dynamics as in the experiment under an external force $F$ applied to each colloid. Although the fluid is not described explicitly, $F$ is to be interpreted as $\eta_{eff}v_d$ where $\eta_{eff}$ and $v_d$ are the effective fluid viscosity and velocity. Particles repel each other with a screened Coulomb interparticle repulsion, Eq. (1), and interact with a triangular-lattice periodic potential

$$W(r) = -(2U_0/9)[3/2 + 2\cos(2\pi r_x/a_L)\cos(2\pi r_y/(\sqrt{3}\,a_L)) + 4\cos(4\pi r_y/(\sqrt{3}\,a_L))]$$

representing the optical lattice corrugation (see Fig. 1B). In addition, a Stokes viscous force $-\eta_{eff}v_i$ acts on each particle $i = 1, \ldots N$, and accounts for the dissipation of the colloids kinetic energy into the thermal bath. Gaussian random forces complete the Langevin thermostat and maintain thermal equilibrium at room temperature $T = T_0 =$ 298 K. We typically simulate $N \simeq 3000$, a particle number much smaller than in experiment but sufficient to extract reliable physical results [1, 2].

The initial configuration is obtained by extracting the colloid coordinates from an experimental snapshot at zero external force and adapting them to a periodic cell compatible with the periodicity of $W(r)$.

The parameters for the colloid-colloid interaction $A$ =12.1 µJµm and $\kappa^{-1}$ = 184 nm have been obtained as a best-fit of the experimental radial-distribution function $g(r)$, weighted by $1/r^2$ in order to stress more the nearest-neighbors peaks. A comparison of simulated and experimental $g(r)$ is provided in Fig. S1.

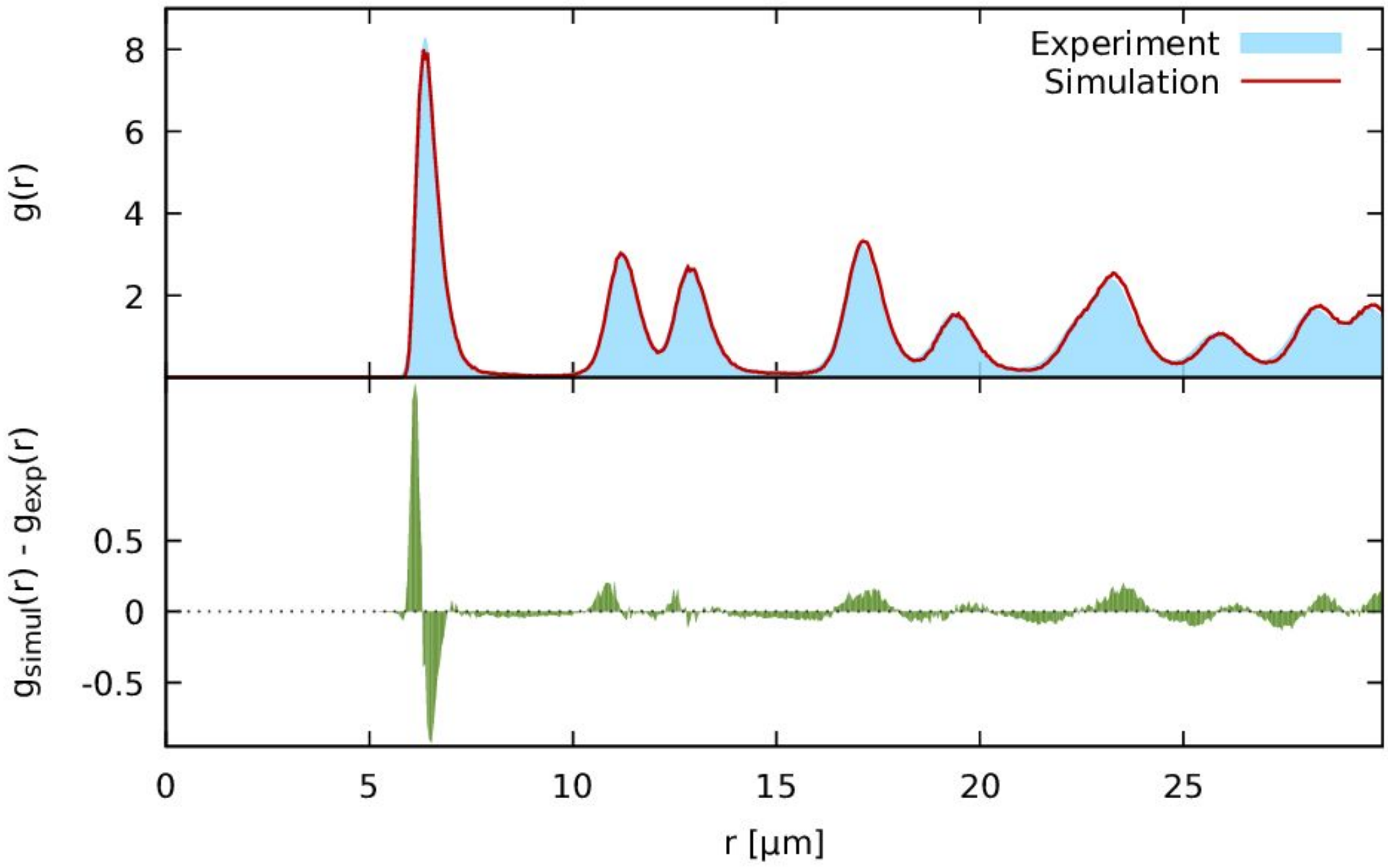

FIG. S1. Comparison of the best-fitting simulated radial distribution function $g_{\mathrm{simul}}(r)$ with the experimental reference $g_{\mathrm{exp}}(r)$. The lower panel details the differences.

## 1.3 Fit of the radial distribution function

To best fit the Yukawa colloid-colloid interaction parameters $A$ and κ we perform the minimization of a suitable error function, evaluated as

$$\chi^2 = \frac{1}{W} \sum_i |g_{simul}(r_i) - g_{exp}(r_i)|^2 \, w(r_i) \, \delta r, \qquad \text{(S1)}$$

where $w(r_i) = 1/r_i^2$ is used to increase the weight of the first peaks, $W$ is the sum of these weights, the mesh interval δr = 0.05 μm, and the number of equally-spaced $r_{\mathrm{i}}$ in Eq. (S1) is approximately 3000. We evaluate $g_{\mathrm{exp}}(r)$ by a thermal average over 600 successive configurations recorded in a static experiment without substrate potential ($U_0$ = 0). Starting from initial parameters $A_0$ and $\kappa_0$ chosen within the experimental estimations, we carry out a simplex minimization of $\chi^2$. Each calculation of the error function requires the following steps:

• run a 300 s-long simulation at room temperature $T = T_0$ in order to first equilibrate the colloid crystal in the presence of the specific colloid-colloid potential defined by the current $A$ and κ values, and then to generate several snapshots for a good sampling of the equilibrium structure and fluctuations;

• evaluate $g_{\mathrm{simul}}(r)$ from those snapshots;

• evaluate $\chi^2$, Eq. (S1), with the newly computed $g_{\mathrm{simul}}(r)$.

The resulting final parameters are $A$ = 12.1 μJμm and $\kappa^{-1}$ = 184 nm. Figure S1 compares the best-fitting $g_{\mathrm{simul}}(r)$ with the reference $g_{\mathrm{exp}}(r)$.

## 1.4 Visualization

In representing snapshots of colloid lattice configurations - from experiment or simulations alike, we highlight the relevant structures by measuring the local "bond" angle $\theta_i$ of each colloid. To do this, for each colloid $i$ we identify all its nearest-neighboring colloids, namely those whose distance from colloid $i$ does not exceed by more than 35% the distance to the colloid $j$ closest to colloid $i$. The number of such neighboring colloids is normally 6; however, due to thermal fluctuations and/or crystalline defects this number can be different, typically 5 or 7. For these "uncoordinated" colloids we have reserved the green color in figures.

For the 6-coordinated colloids, we evaluate the angle $\theta_{ij}$ formed by the vector joining colloid $i$ to each of its 6 neighbors $j$ and the $x$ axis, which is both a crystalline direction of the substrate potential and the direction of the pulling force $F$. Each of these angles $\theta_{ij}$ is evaluated modulo $60^\circ$ within the interval $-30^\circ < \theta_{ij} \le 30^\circ$. The local angle for colloid $i$ is then defined by the average $\theta_i = \frac{1}{6}\sum_{j=1}^{6}\theta_{ij}$. In the snapshots of Figs. 2 and 4 and in the S.I. Movie, we color the colloids according to their $\theta_i$ value: red for $\theta_i > 2.5^\circ$, i.e. locally tilted near the positive Novaco angle; grey for $-2.5^\circ \le \theta_i \le 2.5^\circ$, i.e. essentially not tilted; and blue for $\theta_i < -2.5^\circ$, namely tilted near the negative Novaco angle. The fraction of colloids tilted near the positive or negative Novaco-McTague angle is computed as follows: $f_{tilt} = N(|\theta_i| > 2.5^\circ)/N_6$, where $N_6$ is the number of colloids with 6 nearest neighbors.

In Fig. 4b-c, the arrows indicating the colloids displacement over a 200 s time interval are magnified by a factor 2.45 for better readability. For the same reason, arrows for displacements shorter than 1 μm are not drawn.

## 1.5 Evaluation of the disorder parameter

To evaluate the disorder parameter $\Psi$ we fold all colloid positions back to a single Wigner-Seitz (WS) cell of the corrugation-potential lattice, sketched in Fig. 3d. We then count the number $N_{\mathrm{saddle}}$ of folded particles in the 6 corner triangles of the hexagonal WS cell, corresponding to local corrugation energy exceeding the saddle-point value. The disorder parameter is evaluated as the fraction $\Psi = N_{\mathrm{saddle}}/N$ and averaged over several configurations in order to sample thermal fluctuations.

The initial folding procedure is straightforward for simulation data, where the mutual position of colloids and corrugation potential is known with infinite accuracy. In experiment, the laser-interference pattern generating the corrugation potential fluctuates and drifts slowly in time, with the result that in successive experiments the positions of the corrugation maxima and minima do not remain aligned to the same pixels of the digital microscope recording the colloids. It would practically impossible to monitor at the same time the light pattern and the positions of the colloids. To resolve this difficulty, we take advantage of the colloids being attracted toward the potential wells and pushed away from the repulsive regions above the saddle points (although of course the interparticle forces and thermal fluctuations prevent the colloids to fall all the way to the bottom of the wells). To retrieve the patterns of Fig. 3 we proceed as follows:

- We select a central 150×150 $\mu m^2$ region of the experimental data, where the camera-lens distortion and boundary effects are negligible.
- We select 10 (equally spaced in time) snapshots.
- We define the following 5 parameters:
  1) a global translation along $x$, initially 0 μm
  2) a global translation along $y$, initially 0 μm
  3) an overall rotation angle, initially 0°
  4) the lattice spacing $a_L$, initially estimated at 5.3 μm
  5) a factor for correcting $y/x$ anisotropy effects (e.g. due to camera tilting), initially assumed 1.
- We fold each colloid in the 10 snapshots into the WS cell of the Bravais lattice defined by parameters 4 and 5 above. The colloids will generally be spread out apparently randomly across the WS cell, unless the lattice parameter is correct.
- To fine tune the 5 parameters defined above, we take the data for the greatest available laser intensity, corresponding to corrugation $U_0 = 3.6\ U_c$, where the effect of the corrugation field is strongest, and minimize the sum of the distances of the colloids from the center of the WS cell. An improvement of parameters 3, 4, and 5 leads to a gathering of the colloids into a lump. The adjustment of parameters 1 and 2, moves this lump as near as possible to the center of the WS cell.
- The parameters optimized for the largest corrugation $U_0 = 3.6\ U_c$ are perfect for those data, but lead to randomly translated lumps for the other intensity settings, due to the lateral fluctuations of the light interference pattern. Thus for the lower intensities (smaller $U_0$) we keep the optimized lattice parameter $a_L = 5.38$ μm and anisotropy $y/x = 0.977$ fixed, and re-optimized only over the first three parameters. The resulting optimized angle fluctuates in the -0.6° to -0.2° range, while the translation parameters fluctuate widely.

The resulting well-aligned WS-folded patterns, three examples of which are reported in Fig. 3f-h, are then used to evaluate Ψ as described above.

# 2. Structure factor

We characterize the overall periodic features of the colloidal layer by means of a 2D structure factor evaluated as

$$S(\mathbf{q}) = \frac{1}{N^2} \sum_{i=1}^{N} \sum_{j=1}^{N} e^{\mathbf{i}\mathbf{q}\cdot(\mathbf{r}_i - \mathbf{r}_j)}$$

where $\mathbf{r}_i$ and $\mathbf{r}_j$ are the positions of the particles and the normalization is chosen so that $S(\mathbf{0}) = 1$.

For the experimental data, Fig. S2, the summations extend over the $N \simeq 2730$ particles comprised in the 370x270 $\mu m^2$ viewframe area.

For the simulation, Fig. S3, we include $N \simeq 13700$ particles comprised in a circular portion of the sample of radius 400 μm.

$S(\boldsymbol{q})$ in Figs. S2 and S3 are obtained by averaging over 30 successive configurations.

At weak corrugation amplitude, the structure factor reflects the presence of both $+\theta_{Nov}$ and $-\theta_{Nov}$ in experimental data (Fig. S2a,b), and a single $+\theta_{Nov}$ phase in the simulated data (Fig. S3a,b).

As corrugation is increased (see panels c,e of both Figs. S2 and S3), new peaks arise forming a hexagonal pattern at $|q| \simeq 1.3\ \mu m^{-1}$ and 30°, 90°, 150°, 210°, 270°, and 330° from the *x* axis, representing the substrate corrugation periodicity approached by the colloids within the pinned islands, the latter colored grey in right panels. The weaker peaks forming a hexagonal pattern at $|q| \simeq 0.2\ \mu m^{-1}$, at angles 0°, 60°, 120°, 180°, 240°, and 300° from the *x* axis (Figs. S2e and S3e) reflect the overall average periodicity of the pattern of pinned islands.

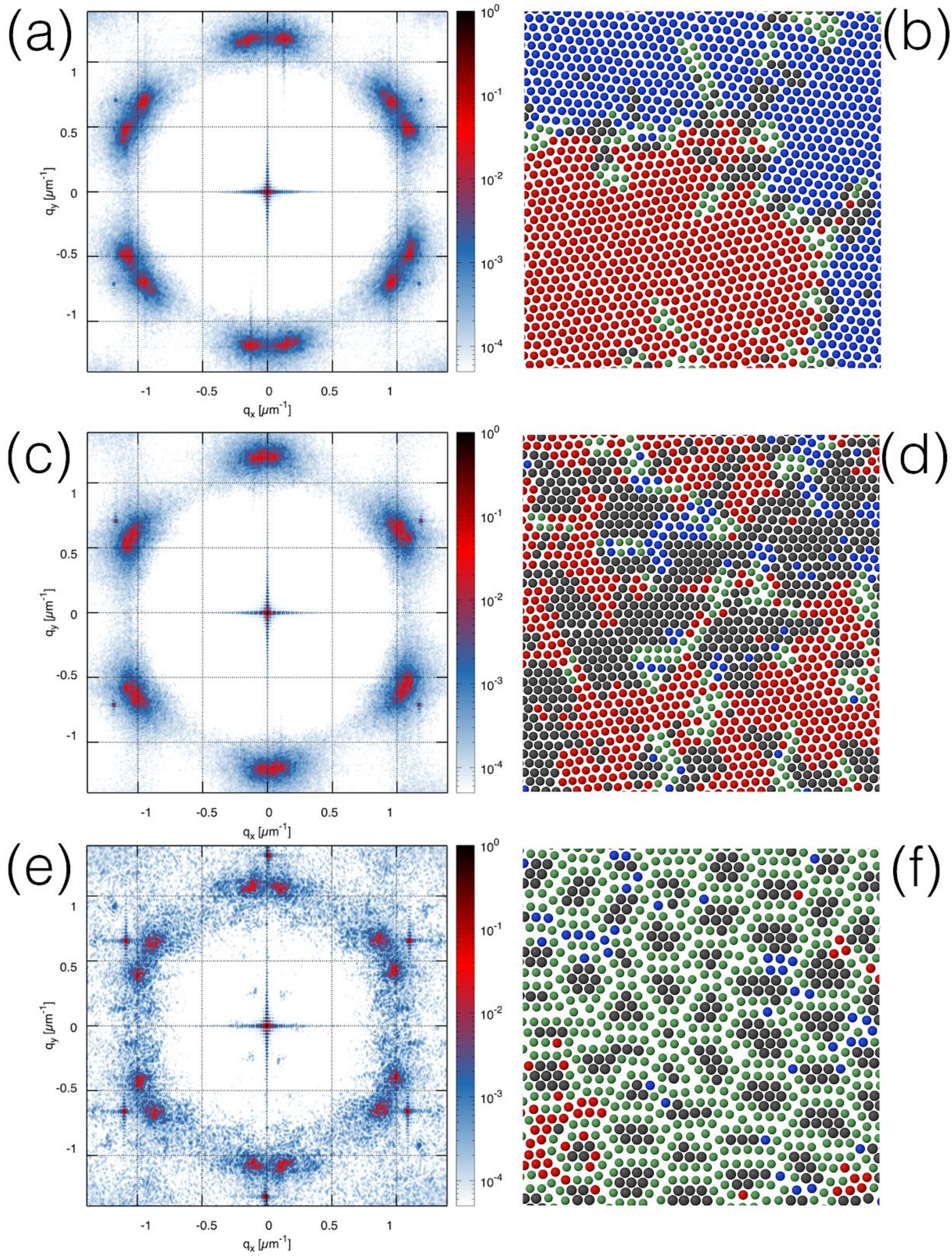

FIG. S2. The structure factor $S(q)$ for the experimental lattice at room temperature $T = T_0$ along with corresponding snapshots of a portion of the system for the following values of the substrate corrugation amplitude: (a,b) $U_0 = 0.63\ U_c$; (c,d) $U_0 = 1.25\ U_c$; (e,f) $U_0 = 11.3\ U_c$. The colloids are colored according to the local bond angle as described in Sec. 1.4 above.

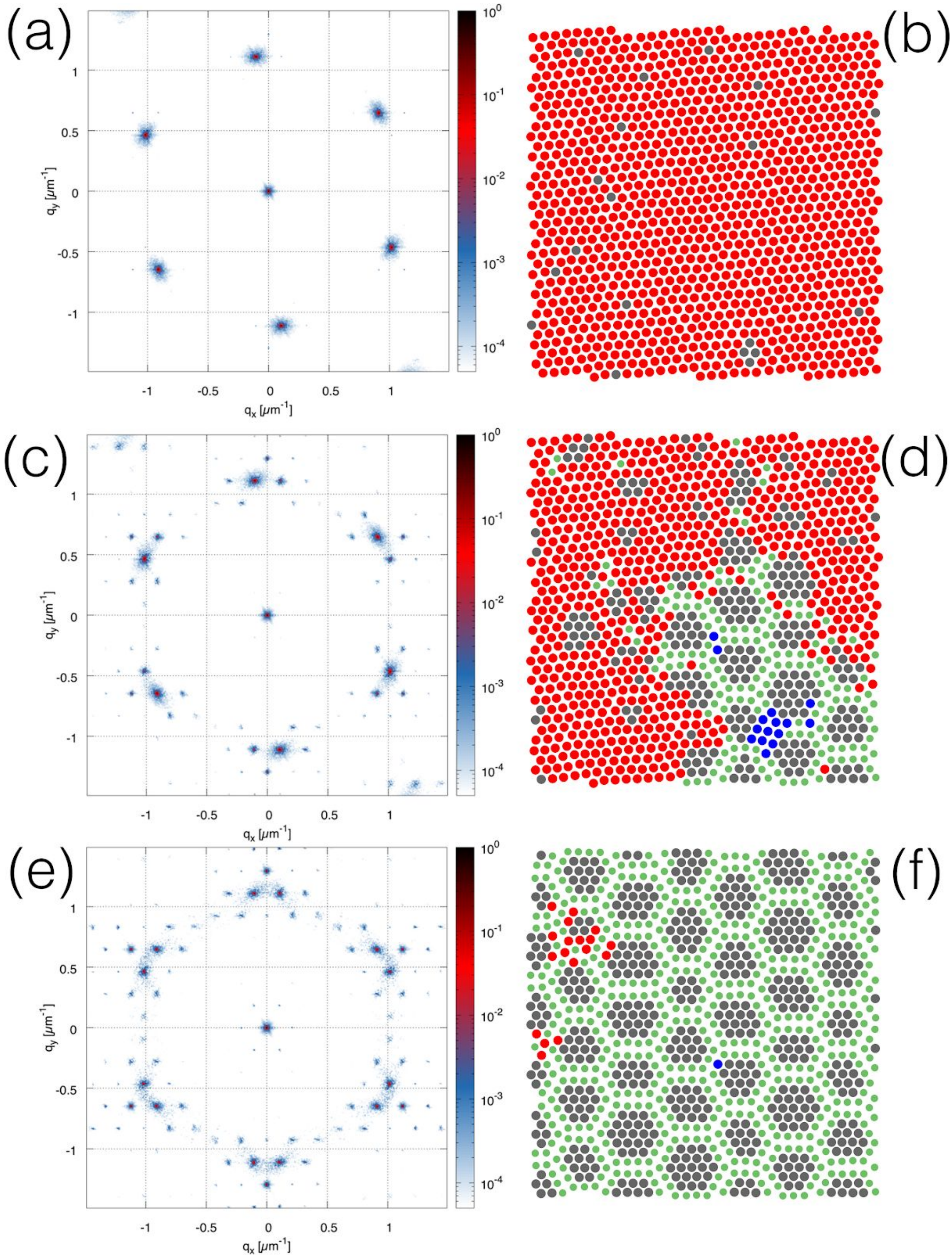

FIG. S3. The structure factor $S(q)$ for the simulated lattice at room temperature $T = T_0$ along with corresponding snapshots of a portion of the system for the following values of the substrate corrugation amplitude: (a,b) $U_0 = 0.21\ U_c$; (c,d) $U_0 = 0.91\ U_c$; (e,f) $U_0 = 3.2\ U_c$. The colloids are colored according to the local bond angle as described in Sec. 1.4 above.